\documentclass[twocolumn,iop,apj]{aastex63}
\usepackage{amsfonts,amsmath,graphicx,natbib,url,hyperref,nicefrac}
\usepackage{amssymb, verbatim, subfigure, paralist, soul, comment}

\hypersetup{colorlinks=true, urlcolor=blue, citecolor=blue}

\usepackage{color}
\newcommand{\add}[1]{\textcolor{black}{#1}}

\newcommand{\sprout}{\texttt{Sprout}}

\usepackage{comment}

\shorttitle{Hydrodynamics on expanding mesh} 
\shortauthors{Mandal \& Duffell}

\begin{document}

\title{SPROUT: A moving mesh hydro code using a uniformly expanding Cartesian grid}

\author{Soham Mandal}
\affiliation{Department of Physics and Astronomy, Purdue University, 525 Northwestern Avenue, West Lafayette, IN 47907, USA}

\author{Paul C. Duffell}
\affiliation{Department of Physics and Astronomy, Purdue University, 525 Northwestern Avenue, West Lafayette, IN 47907, USA}

\begin{abstract}

We present the publicly available moving-mesh hydrodynamics code \sprout. \sprout\;solves the equations of ideal hydrodynamics on an expanding Cartesian mesh. The expanding mesh can follow fluid outflows for several orders of magnitude with very little numerical diffusion, thereby capturing shocks and fine structures accurately. Following the bulk flow accurately also allows for longer timesteps in general. This makes \sprout\;particularly suitable for studying expanding outflows such as supernova remnants and active galactic nuclei. Relative to other moving mesh codes, the simple mesh structure in \sprout\;is also convenient for implementing additional physics or algorithms. Many code tests are performed to test the accuracy and performance of the numerical scheme.

\end{abstract}


\section{Introduction} \label{sec:intro}

Many astrophysical phenomena involve fluid flows that expand over several orders of magnitude in size and time. Computational studies of such outflows require numerical schemes that can handle large dynamic ranges accurately. For instance, numerical studies of supernova remnants often require modeling of the stellar ejecta starting from shock breakout (or before) and evolving the solution upto tens of thousands of years \citep{Hammer+2010ApJ,Wongwathanarat+2015,Wongwathanarat+2017ApJ}. This can cause an expansion of more than three orders of magnitude. An even more dramatic example is the jet from a supermassive black hole, which can grow up to many kiloparsecs (kpcs). They have been modeled for a very wide range of physical scales, e.g., jet launching from BHs \citep{DeVilliers+2003ApJ,McKinney+2009MNRAS}, parsec and kpc scale jets \citep{Komissarov1999MNRAS,Porth+2011ApJ}. Even larger length scales (hundreds of megaparsecs) are required for numerical studies of cosmology and large-scale structure formation \citep{Genel+2014MNRAS,Schaye+2015MNRAS}, which also need to resolve galaxy scale features reasonbly well.

Lagrangian methods such as Smoothed Particle Hydrodynamics \citep[SPH;][]{Gingold+1977,Springel2005,Springel2010ARA&A,Price+2018} are a natural choice for studying fluid flows that span many orders of magnitude in size. The naturally adaptive resolution of these methods allows great dynamic ranges in spatial resolution but as well as density. They do not come with the advantages of shock-capturing and high-order convergence typical of Eulerian grid-based codes \citep{Leer1977,Woodward+1984,Stone+2008ApJS}. Eulerian methods often handle the problem of high dynamic ranges using adaptive mesh refinement \citep[AMR;][]{Fryxell+2000ApJS,Ono+2013ApJ,Brian+2014ApJS,Kostic2019SerAJ} to derefine the mesh into a new, larger mesh as the outflow is about to cross the boundary. Another solution is to use grid-based methods to solve the equations of hydrodynamics in a non-inertial frame that is comoving with the bulk of the fluid \citep[e.g.,][]{Poludnenko+2007,Brant+2012ApJ}.

An alternative direction that has received much attention over the years is development of numerical methods that retain the advantages of both Lagrangian and Eulerian paradigms. A notable example is the moving mesh (MM) approach \citep{Springel2010MNRAS,Duffell+2011ApJS,Gaburov+2012ApJ}. MM codes solve hydro equations using finite volume methods (FVM) on meshes (often unstructured) that restructure themselves following the fluid flow. Mesh free methods \citep[MFM and MFV;][]{Hopkins2015MNRAS} form a class of closely-related approaches to moving mesh methods. This also consists of a finite volume approach to discretization of fluid equations, but the key difference is that volume is distributed among particles according to a weighting function, as opposed to discrete zone boundaries in moving mesh schemes. Even though such methods can accurately follow fluid flow of any nature and over large length scales, it is often advantageous to restrict the mesh motion roughly to the bulk flow velocity, provided this is known before runtime. This idea marked the advent of MM codes like DISCO \citep{DISCO} and JET \citep{JET}, which are tailored to studying azimuthal flows (like discs) and radial outflows like jets or supernovae, respectively. JET can handle flows with very high dynamic ranges by allowing its boundaries to move with time.

The Euler-Lagrange hybrid hydrodynamics codes have proven to be reliable and useful in solving numerous astrophysical problems that involve great dynamic ranges. Their success motivated us to develop a distinct numerical scheme for studying expanding flows using the restricted mesh motion methodology. This work describes our novel moving-mesh hydro code, \sprout. \sprout\, is designed to expand its boundaries over time. This has been implemented before in numerical schemes such as ZEUS \citep{Stone+1992ApJS,Jun+1996ApJ} or the Virginia Hydrodynamics 1 code, or VH-1 \citep{VH1}. ZEUS, for instance, uses an additional transport step for fluid advection, which may be modified to take into account the grid velocity. VH-1 utilizes a Lagrangian remapping step to remap the hydro variables into a new grid after each timestep. \sprout\, employs a different strategy for expanding its domain, which is more similar to JET. But unlike JET, the mesh motion does not follow in detail the local fluid velocity but is restricted only to expand (or contract) as a whole, preserving its structure. Although this dilutes the usual advantages offered by moving mesh codes like drastic reduction of numerical diffusion or allowing for significantly longer timesteps, we are left with an efficient Cartesian scheme that can handle high dynamic ranges and allows for simple implementation of complex algorithms and additional physics like magnetic fields. The choice of geometry was made to avoid issues that conventional curvilinear grids often face, such as geometrical singularities and very small signal crossing time near the poles.

\sprout\, has been successfully employed to study the properties of Rayleigh-Taylor instabilities developing in young supernova remnants \citep{Mandal+2023arXiv}. The numerical scheme of \sprout\, is discussed in detail in Section \ref{sec:numerical}. Section \ref{sec:tests} describes the numerous tests performed to examine stability, accuracy and performance of our numerical scheme. The results are summarized in Section \ref{sec:summary}.

\section{Numerical method} \label{sec:numerical}

\subsection{Field equations} \label{subsec:equations}

\sprout $\;$is designed to solve Euler's equations in the conservation law form as follows:

\begin{equation}
\label{eq:conservation_law}
    \partial_t u + \nabla \cdot  \mathbf{F}  = S
\end{equation}

Thus, Euler's equations are expressed as:

\begin{equation}
\label{eq:euler}
    \begin{gathered}
        \partial_t(\rho) + \nabla \cdot ( \rho \mathbf{v} ) = S_D \\
        \partial_t( \rho \mathbf{v} ) + \nabla \cdot ( \rho \mathbf{vv} + P \overleftrightarrow{I} ) = S_M \\
        \partial_t\left( \frac{1}{2}\rho v^2 + \epsilon \right) + \nabla \cdot \left( \left( \frac{1}{2}\rho v^2 + \epsilon + P \right)\mathbf{v} \right) = S_E,
    \end{gathered}
\end{equation}

where $\rho$, $\mathbf{v}$, $P$ and $\epsilon$ are the primitive variables and stand for density, velocity, pressure and internal energy respectively. The terms $S_D$, $S_M$ and $S_E$ are source terms for mass, momentum and energy respectively and allow us to introduce external gravity and/or engine driven outflows such as jets. $P$ is related to $\epsilon$ via an adiabatic equation of state:

\begin{equation}
    P = (\gamma-1)\epsilon,
\end{equation}

where $\gamma$ is the adiabatic index.

\begin{figure*}
\gridline{\fig{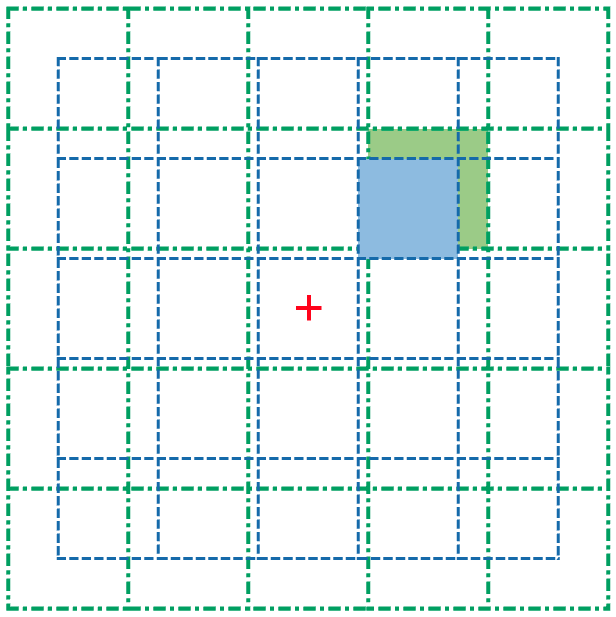}{0.4\textwidth}{(a)}
          \fig{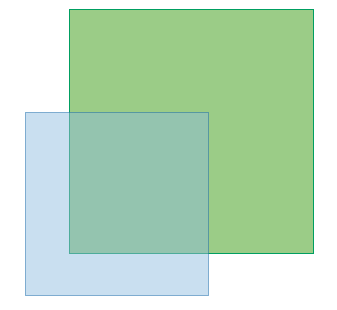}{0.4\textwidth}{(b)}
          }
\centering
\caption{(\textit{Left}) Two instances of an expanding 2D grid, with the later instance shown in green. The red cross marks the center of expansion. (\textit{Right}) Zoomed-in view of one zone, which illustrates how the zone faces move in general. This also shows that the zone areas and volumes in Equation \ref{eq:evolution_moving} have to be treated as functions of time while evaluating the integrals.
\label{fig:grid}}
\end{figure*}

\subsection{Mesh Description} \label{sec:mesh}

\sprout $\;$uses a finite volume method to discretize the field equations over the computational domain. The domain is divided into a finite number of cells or zones using a uniform Cartesian mesh. This mesh can be allowed to expand self-similarly in time, that is, the mesh dimensions can change keeping all aspect ratios fixed. This is depicted in Fig~\ref{fig:grid} for a 2D mesh. It can be shown that the mesh can transform self-similarly for any timestep of span $\Delta t$, if and only if any point on the mesh with a position vector $\mathbf{r}$ before the transformation can be located by a new position vector as follows:

\begin{equation}
\label{eq:mesh_motion}
    \mathbf{r'} = \mathbf{r} + H \left( \mathbf{r}-\mathbf{r_0} \right) \Delta t,
\end{equation}

where $\mathbf{r_0}$ is the reference position vector and H is a (time-dependent) constant that essentially sets the rate of expansion of the mesh, similar to the Hubble constant. H should stay uniform over the grid, but may have a different value for each timestep. (\ref{eq:mesh_motion}) provides the prescription for velocity of all computational zone boundaries or ``faces". The other important consequence of mesh motion is on the separation between two points fixed relative to the mesh. Any such separation $\Delta x$ is transformed after one timestep of span $\Delta t$ according to:

\begin{equation}
\label{eq:separation}
    \Delta x' = \Delta x \left( 1 + H \Delta t \right)
\end{equation}

This can easily be seen by applying (\ref{eq:mesh_motion}) to two distinct points.

\subsection{Modified form of integral equations} \label{sec:flux_update}

We are now ready to discretize the field equations. Equation (\ref{eq:conservation_law}) can be integrated over the volume of a computational zone to give:

\begin{equation}
\label{eq:integral_form}
    \int \partial_t u dV + \oint d\mathbf{A} \cdot \mathbf{F}  = \int SdV,
\end{equation}

where the second term is obtained by using Gauss' theorem on the divergence of the flux. As long as the zone boundaries do not move, that is, the mesh is static, the time derivative on the first term can be taken out of the integral, allowing us to write:

\begin{equation}
\label{eq:mass_integral_static}
    \int (\partial_t u) dV = \partial_t \left(\int u dV\right)
\end{equation}

Let's consider the $i$-th zone of the computational domain during timestep number $n$. We can then integrate Equation \ref{eq:integral_form} over the timestep $\Delta t$ and the zone volume $\Delta V$ to write

\begin{equation}
\label{eq:evolution_static}
    M_i^{n+1} = M_i^n - \int dt \oint d\mathbf{A} \cdot \mathbf{F} + \int dt \int S dV, 
\end{equation}

\add{where $M_i^n=\int u_i (t_n) dV$ and stands for the total amount of conserved quantity $u$ in the $i$-th zone at the $n$-th timestep $t_n$}. If there is mesh motion, (\ref{eq:mass_integral_static}) can be generalized using Reynold's transport theorem \citep[see also][]{DISCO}:

\begin{equation}
    \int (\partial_t u) dV = \partial_t \left(\int u dV\right) - \oint u\mathbf{w} \cdot d\mathbf{A},
\end{equation}

\add{where $\mathbf{w}$ stands for the velocity of the corresponding zone face.} Thus Equation \ref{eq:evolution_static} is modified as follows:

\begin{equation}
\label{eq:evolution_moving}
    M_i^{n+1} = M_i^n - \int dt \oint d\mathbf{A} \cdot \left(\mathbf{F}-\mathbf{w}u\right) + \int dt \int S dV 
\end{equation}

Up to this point, we have not evaluated the flux or source term integrals explicitly and therefore the evolution equation remains exact. We now note that the area of a zone face or the volume of a zone is a function of time if the mesh moves. This dependence can be explicitly written by using Equation (\ref{eq:separation}):

\begin{equation}
    \begin{split}
        dA(t+\tau) &= dA(t)\left(1+H\tau\right)^{D-1}, \\
        dV(t+\tau) &= dV(t)\left(1+H\tau\right)^{D}, \\
    \end{split}
\end{equation}

where $D=1,2,3$ is the number of dimensions. To evaluate the integrals in (\ref{eq:evolution_moving}), we set $t=0$ and $\tau=\Delta t$ in the above expressions and perform integrals of the form $\int dA(\tau) d\tau$. Note that the fluxes and the source terms are assumed to be constant over the timestep, which should be more erroneous with mesh motion turned on. However, as can be seen in the code tests, our scheme is able to maintain a second-order accuracy with an expanding mesh. Also, $H$ is not treated as a function of time, since $H$ is taken to be a constant for each timestep (but not necessarily over multiple timesteps). This gives us the most generalized form of our evolution equation:

\begin{multline}
\label{eq:evolution_final}
    M_i^{n+1} = M_i^n - C_F\,\Delta t \sum_{face\,j}\Delta\mathbf{A_j} \cdot \left(\mathbf{F_{ij}}-\mathbf{w_j}u\right) \\ + C_S\,\Delta t\,\Delta V S_i,
\end{multline}

where we introduce correction factors for the flux and source terms arising due to change in the zone volume and face area over a timestep, given by:

\begin{equation}
    \begin{split}
        C_F &= \frac{(1+H(t^n)\Delta t)^{D}}{D H(t^n)\Delta t} \\
        C_S &= \frac{(1+H(t^n)\Delta t)^{D+1}}{(D+1) H(t^n)\Delta t} \\
    \end{split}
\end{equation}

Equation (\ref{eq:evolution_final}) is exact as long as $\mathbf{F_{ij}}$ is interpreted as the time-averaged and area-averaged flux in zone $i$ through face $j$ and $S_i$ is interpreted to be the time-averaged and volume-averaged source term in zone $i$.

\subsection{Riemann solver}

Following the Godunov formulation, the next step to solving Equation (\ref{eq:evolution_final}) is to specify a prescription for the flux through any given face, given the conserved quantities are known over the domain. This is done by using a Riemann solver.

Let's assume we require a solution to a hyperbolic conservation law equation of the form (\ref{eq:conservation_law}) in 1 dimension with the following initial condition:

\begin{equation}
u\left(x,\,t=0\right)=
\begin{cases}
  u_L, & x<0 \\
  u_R, & x>0
\end{cases}
\end{equation}

A Riemann solver takes this initial condition and calculates the value of the conserved quantity, and the flux at a later time t. Thus, it provides the interface ($x=0$) flux $F_{\*}$:

\begin{equation}
    F_{*} = F\left(x=0,\,t\right)
\end{equation}

However, we wish to know the flux at an interface which is not static but moves with some known velocity $w$. Thus, we evaluate the flux and the conserved quantity along the characteristic $x=wt$ (cf. Figure 3 of \cite{DISCO}):

\begin{equation}
    F_{*} = F\left(x=wt,\,t\right)
\end{equation}

We use the HLLC approximate Riemann solver \citep{Toro+1994}, which is very useful for maintaining contact discontinuities with high precision. The solution to the Riemann problem thus enables us to apply Equation \ref{eq:evolution_final} to update the hydro variables in every zone for a known timestep $\Delta t$.

\subsection{Second order interpolation}

So far, we have assumed the primitive variables to be uniform over any given zone. We need a better approximation particularly while specifying initial conditions to the Riemann solver to get higher order spatial accuracy. \sprout\, achieves second-order spatial accuracy by linearly interpolating the primitive variables in between zones. \add{Let's consider the variable $W$ is being interpolated in the $x$-direction. Then, its value between the $i$-th and the $(i+1)$-th zone will be}:

\begin{equation}
    W_{i\pm\frac{1}{2}} = W_i \pm \frac{1}{2}\Delta x\cdot\partial_x W_i,
\end{equation}

\add{where $\Delta x$ is the zone width in the $x$-direction, and $\partial_x W_i$ is the $x$-component of the gradient of $W$ in the neighborhood of the $i$-th zone.} The value of the gradient depends on the left, right or centered slopes $S_L$, $S_R$ and $S_C$, which are defined as follows:

\begin{equation}
    \begin{split}
        S_L &= (W_i-W_{i-1})/\Delta x \\
        S_R &= (W_{i+1}-W_i)/\Delta x \\
        S_C &= (W_{i+1}-W_{i-1})/(2\Delta x)
    \end{split}
\end{equation}

A slope limiter is required to choose the gradient value. This helps to avoid spurious oscillations in the neighborhood of discontinuities of the primitive variable. In this work, we use the minmod slope limiter:

\begin{equation}
    \nabla_x W = \text{minmod}(\theta_{plm}S_L, \theta_{plm}S_R, S_C) 
\end{equation}

The parameter $\theta_{plm}$ may vary between $1-2$ and can be adjusted to gain better second-order correction while ensuring stability. The minmod function is defined as:

\begin{equation}
\text{minmod}\left(x,y,z\right)=
\begin{cases}
  \text{min}(x,y,z), & x,y,z>0 \\
  \text{max}(x,y,z), & x,y,z<0 \\
  0, & \text{otherwise}
\end{cases}
\end{equation}

The minmod slope limiter assures there are no false discontinuties near extrema of the variables. The numerical scheme reverts to first-order in the neighborhood of zones with extremum values.

\subsection{Time evolution}

The time evolution of the system is described by Equation (\ref{eq:evolution_final}). It can be re-expressed as:

\begin{equation}
\label{eq:evolution_feuler}
    M_i^{n+1} = M_i^n + \Delta t\,L_i\left(\,\{\text{state}\,n\}\,\right),
\end{equation}

where $L_i$ is interpreted as the time-evolution operator at zone $i$ acting on the fluid variables at timestep $n$. This equation is completely specified now with the exception of the timestep $\Delta t$. It is set by imposing the Courant–Friedrichs–Lewy condition in the following way:

\begin{equation}
    \Delta t = C_{\text{CFL}} t_{\text{cross}}, 
\end{equation}

where $C_{\text{CFL}}<1$ and $t_{\text{cross}}$ is the minimum amongst the signal-crossing times of all zones over all dimensions $j$:

\begin{equation}
    t_{\text{cross}} = \text{min}\left( \frac{\Delta x_j}{c_s+\left|v_j-w_j\right|} \right),
\end{equation}

where $c_s$ denotes the speed of sound in a given zone. \add{The quantities $\Delta x_j$, $v_j$ and $w_j$ stand for the width, fluid velocity and mesh velocity of that zone in the $j$-th direction respectively. Note that mesh motion enables the cell velocities $w_j$ to be approximately equal to the bulk component of the fluid velocity $v_j$, allowing for larger timesteps.} Comparing Equation \ref{eq:evolution_feuler} to an evolution equation of the form

\begin{equation*}
    y_{n+1} = y_n + \Delta t \frac{dy}{dt}\bigg|_{t=t_n},
\end{equation*}

we see that our time-evolution operation is only accurate to first-order in time. \sprout\, is capable of achieving higher order accuracy in time by use of a low-storage method-of-lines integration algorithm due to \cite{shu1988efficient}. Second order accuracy is obtained using the Runge-Kutta (RK2) scheme. This requires evaluation of an intermediate state $M^{(1)}$ as follows:

\begin{align}
    M_i^{(1)} &= M_i^n + \Delta t\,L_i\left(\,\{\text{state}\,n\}\,\right), \\
    M_i^{n+1} &= \frac{1}{2}(M_i^n+M_i^{(1)}) + \frac{1}{2}\Delta t\,L_i\left(\,\{\text{state}\,(1)\}\,\right).
\end{align}

The RK3 scheme, which is the third-order accurate analogue of the above, is also available in \sprout. This algorithm requires another intermediate state $M^{(2)}$:

\begin{align}
    M_i^{(1)} &= M_i^n + \Delta t\,L_i\left(\,\{\text{state}\,n\}\,\right), \\
    M_i^{(2)} &= \frac{3}{4}M_i^n + \frac{1}{4}M_i^{(1)} + \frac{1}{4}\Delta t\,L_i\left(\,\{\text{state}\,(1)\}\,\right), \\
    M_i^{n+1} &= \frac{1}{3}M_i^n+\frac{2}{3}M_i^{(2)} + \frac{2}{3}\Delta t\,L_i\left(\,\{\text{state}\,(2)\}\,\right). 
\end{align}

In practice, the third-order accurate timestepping is sometimes found to be useful for transporting data between zones that share a common corner, reducing odd-even decoupling \citep[e.g.][]{Quirk1994IJNMF} and alleviating the effect of certain numerical instabilities.

\subsection{Parallelization}

\sprout\, utilizes the Message Passing Interface (MPI) to enable parallel operation on a large number of processors. The computational domain is subdivided along all dimensions of operation. A 3D domain, for example, is broken up into cuboids. Boundary data is communicated amongst CPUs along each dimension.

\section{Test problems} \label{sec:tests}

We perform a large number of tests to examine accuracy and convergence of \sprout's numerical scheme. Many commonly-used tests for hydrodynamics were performed to ensure correct order of convergence and accurate shock-capturing. Further tests, relevant in astrophysical context, were carried out to examine \sprout's capability to resolve fine features and minimize numerical diffusion when the expanding mesh is utilized. All tests were conducted with an adiabatic index ($\gamma$) of $5/3$ and $C_{\text{CFL}}=0.5$, unless mentioned otherwise. \add{A few of these test results were compared to similar tests performed for other codes. In these cases, we define N as the number of cells or resolution elements per dimension for all codes to enable ease of comparison of results.}

\begin{figure}
\centering
\fig{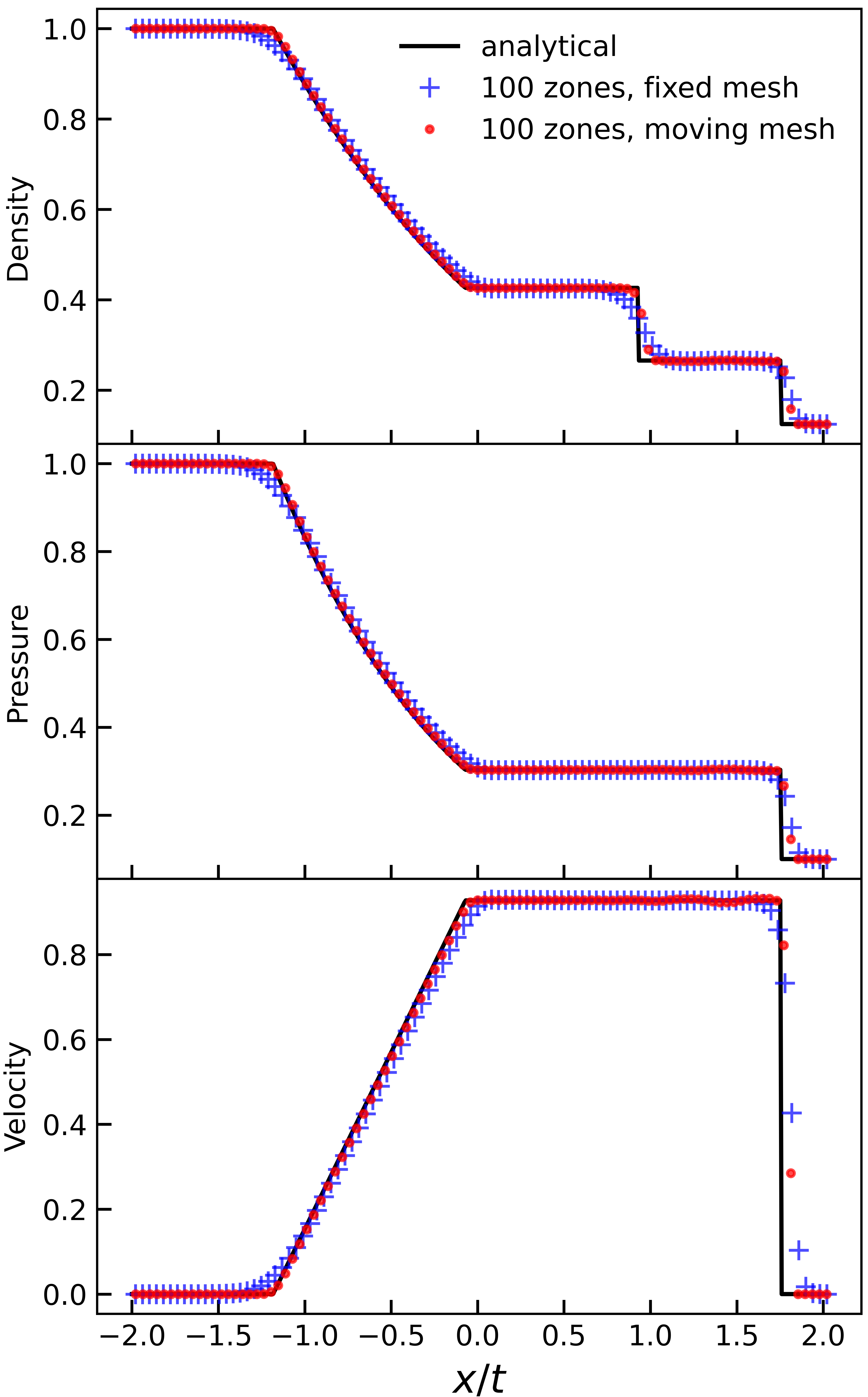}{0.45\textwidth}{}
\caption{Density plotted against distance from the shock scaled with time for the 1D shock tube test, at $t=100$.  The moving mesh results (dashed lines) and the fixed mesh result (plus symbols) are compared to the analytical solution (solid line). The moving mesh captures the shocks and contact discontinuities more accurately when compared to the fixed mesh, and in addition preserves self-similarity after a expansion by two orders of magnitude. In both cases, 100 computational zones are used.
\label{fig:Sod_all}}
\end{figure}

\subsection{1D shock tube} \label{sec:Sod}

The 1D Sod shock tube test is a simple test used to test shock-capturing capability of numerical schemes. The domain is defined on $x \in [-L,L]$. The following initial conditions are used, along with an adiabatic index $\gamma=1.4$:

\begin{equation}
   \begin{split}
       \rho &= 1.0,\,P = 1.0;\;\;x\leqslant0, \\
       \rho &= 0.125,\,P = 0.1;\;\;x>0, \\
       v &= 0. \\
   \end{split}
\end{equation}

This test is performed with both the fixed mesh and the expanding mesh configuration. In both cases, 100 computational zones are used and calculations are made till the time $t=100$. We choose $L=200$ for the former, while the expanding mesh calculations are initialized with $L=2$. The center of expansion is located at $x=0$. The mesh expansion is chosen to follow the shocks originating from the initial point of discontinuity and moving in either direction. Fig \ref{fig:Sod_all} shows the density, pressure and velocity for both cases plotted against $x/t$. This demonstrates \sprout's ability to capture shocks on an expanding mesh over several orders of magnitude in time. The expanding mesh configuration preserves the self-similarity of the setup, which arise from the scale-free initial conditions.

\begin{figure}
\centering
\fig{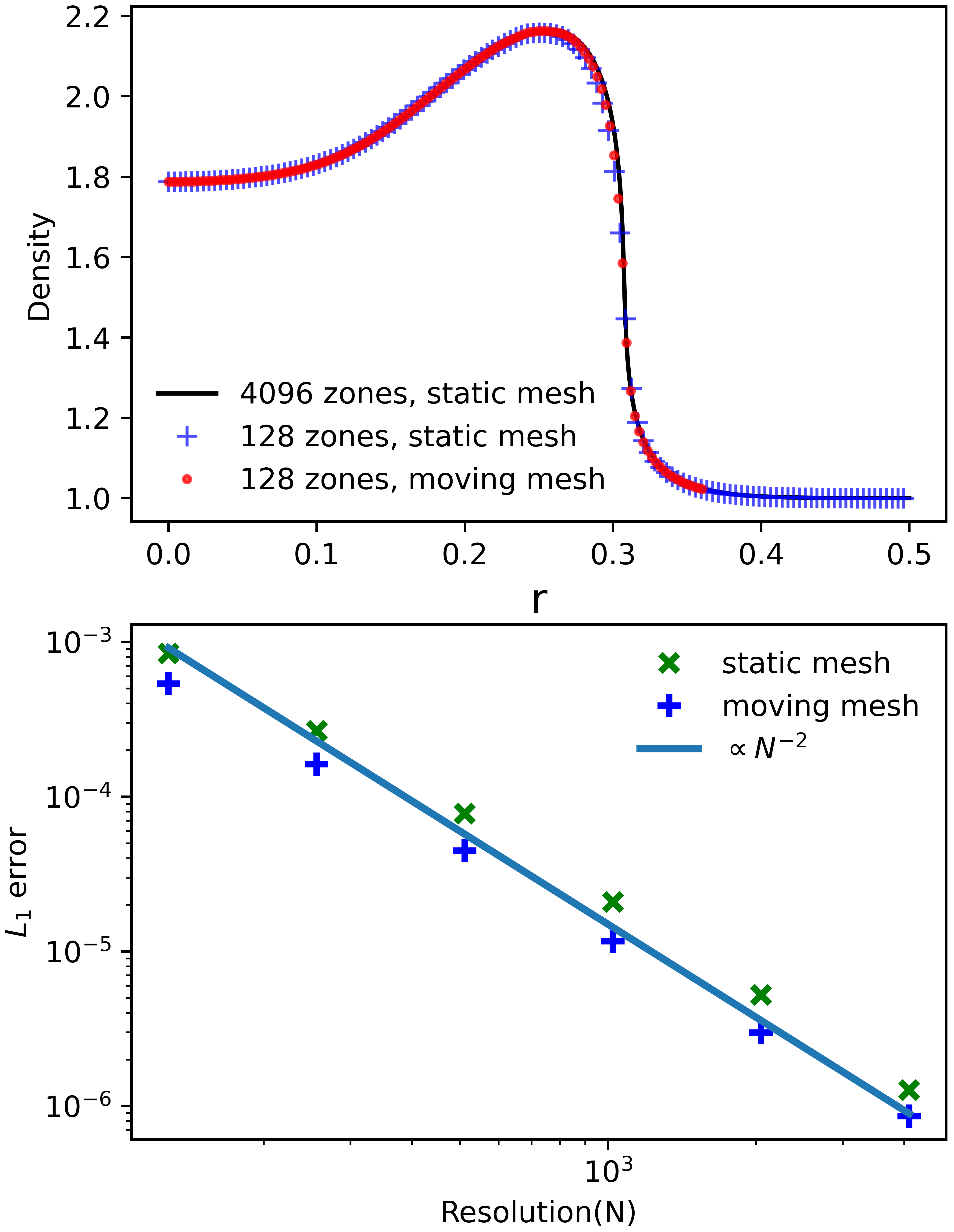}{0.45\textwidth}{}
\caption{Convergence of the 1D isentropic pulse test. (\textit{Top}) Density vs distance at t=0.1 obtained using both the static and expanding mesh, compared against a high-resolution calculation (in blue) with mesh motion turned off. (\textit{Bottom}) $L_1$ error as a function of resolution (Equation \ref{eq:l1}).
\label{fig:1Dpulse}}
\end{figure}

\subsection{1D isentropic pulse}

The isentropic pulse is used as a simple convergence test for many numerical schemes. We calculate errors by measuring departure from entropy conservation, since \sprout\, does not conserve entropy explicitly. The initial conditions are as follows:

\begin{equation}
\label{eq:isentropic_pulse}
   \begin{split}
       \rho &= 1 + 3e^{-80x^2}, \\
       P &= \rho^{\gamma}, \\
       v_x &= 0. \\
   \end{split}
\end{equation}

The domain is defined in $0<x<0.5$. Reflective boundary conditions (BCs) are imposed on the left hand side boundary of the domain, while the right hand side implements Dirichlet BCs. The pulse is allowed to expand till $t=0.1$. As long as no shock is formed, entropy conservation, and therefore the relation $P=\rho^{\gamma}$ holds. Thus we can obtain an error by calculating the following quantity:

\begin{equation}
\label{eq:l1}
    L_1 = \frac{\int \left|P/\rho^{\gamma}-1.0\right| dV}{\int dV}
\end{equation}

We repeat this test with a moving mesh. We start with a smaller domain ($0<x<0.3$) and expand the mesh following the expanding pulse. This essentially allows us to better resolve the fluid at all times compared to a static mesh with the same number of zones. The upper panel of Figure \ref{fig:1Dpulse} shows the density profile of the pulse at $t=0.1$ for both the static mesh and the moving mesh calculations compared to a high-resolution run. Qualitative agreement is found for both cases. The moving mesh calculation seems to be marginally more accurate. The L1 errors in the lower panel show that we indeed get $\sim40\%$ less error for the moving mesh case. Second-order convergence is found for both mesh configurations.

\begin{figure}
\centering
\fig{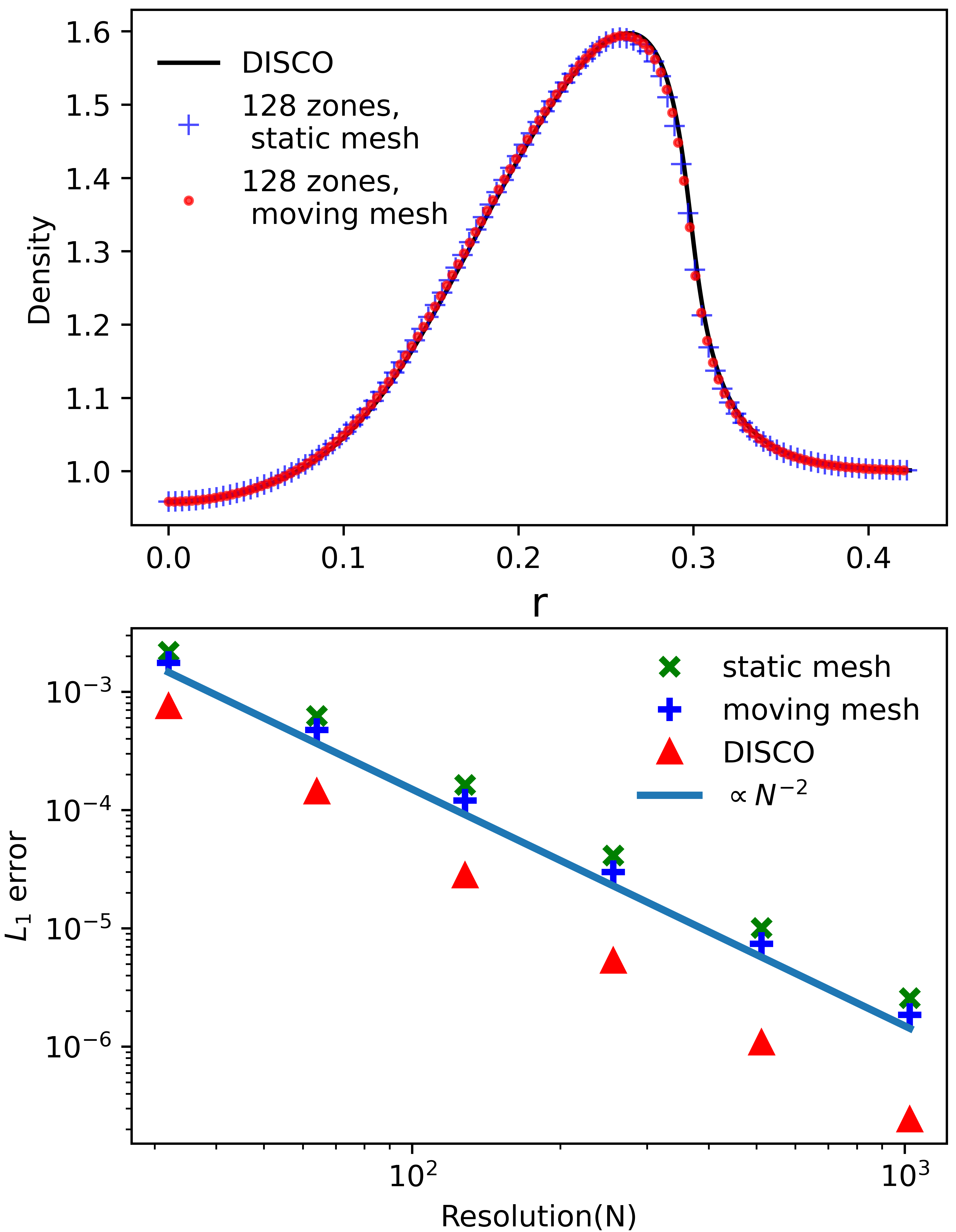}{0.45\textwidth}{}
\caption{Convergence of the 2D isentropic pulse test. (\textit{Top}) Angle averaged density vs radial distance at t=0.1 (with static mesh), compared against a high-resolution 1D solution obtained by the DISCO code (in blue) in cylindrical coordinates. (\textit{Bottom}) $L_1$ error as a function of resolution (Equation \ref{eq:l1}), along with the same for the cylindrical isentropic pulse test done for DISCO. \add{N stands for number of zones per dimension for both codes.}
\label{fig:2Dpulse}}
\end{figure}

\subsection{2D isentropic pulse}

The 2D version of the isentropic pulse test serves as a multidimensional convergence test. The initial conditions remain the same as in Equation (\ref{eq:isentropic_pulse}) except for the density, which is now given by:

\begin{equation*}
    \rho = 1 + 3e^{-80(x^2+y^2)}.
\end{equation*}

We use a static mesh for this test with the domain defined on $(x,y) \in [0,1]\times[0,1]$. The boundary conditions remain similar to the 1D isentropic pulse problem (reflective BCs for the $x=0$ and $y=0$ boundaries, Dirichlet BCs for the rest). This setup has cylindrical symmetry. Thus it is essentially the same test as the cylindrical isentropic pulse test for DISCO (section 3.1.2 of \cite{DISCO}). This allows us to compare our L1 error against DISCO's. The match between DISCO's and \sprout's solutions also serve as a sanity check for our BCs. We compute angle-averaged density as a function of radius (or distance from the origin) at $t=0.1$ and compare it with DISCO, again demonstrating second-order convergence, as shown in Figure \ref{fig:2Dpulse}.

\begin{figure}
\centering
\fig{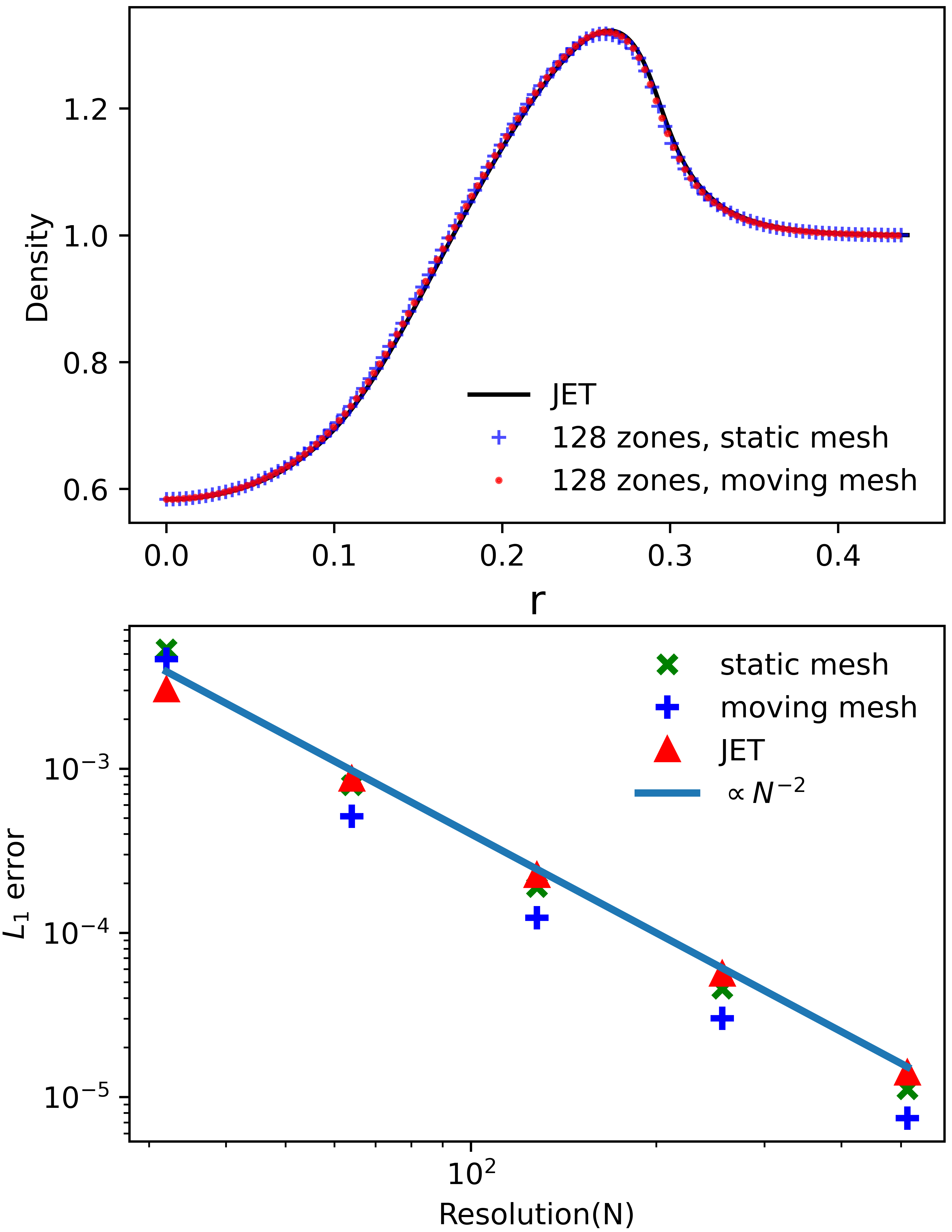}{0.45\textwidth}{}
\caption{Convergence of the 3D isentropic pulse test. (\textit{Top}) Angle averaged density vs radial distance at t=0.1 for both the static and the expanding mesh, compared against a high-resolution solution obtained by the JET code for a 1D isnetropic pulse in spherical coordinates (in blue). (\textit{Bottom}) $L_1$ error as a function of resolution (Equation \ref{eq:l1}), along with the same for the spherical isentropic pulse test done for JET. \add{N stands for number of zones per dimension for both codes. We note that the exact definitions of $L_1$ for the two codes are not consistent, hence it makes more sense to compare the convergence rate than actual error values.}
\label{fig:3Dpulse}}
\end{figure}

\begin{figure*}
\centering
\fig{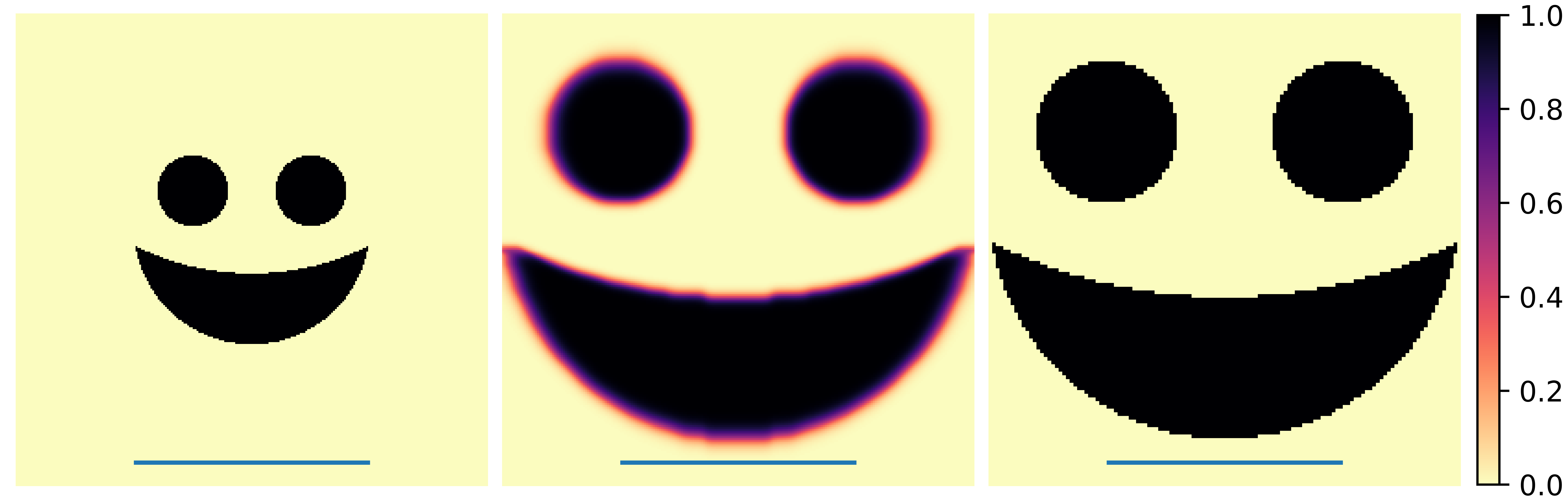}{0.8\textwidth}{}
\caption{Passive scalar evolution for homologous outflow. (\textit{Left}) Passive scalar map at the initial time ($t=1$). (\textit{Middle}) Passive scalar map at $t=2$, with mesh motion turned off. (\textit{Right}) Same as middle, but with expanding mesh, following the homologous expansion. The latter shows no measurable numerical diffusion, as opposed to the static mesh run. The blue line in each panel has a length of unity in code units.
\label{fig:homologous}}
\end{figure*}

\subsection{3D isentropic pulse} \label{sec:3dpulse}

The 3D isentropic pulse test is very similar to the 1D and 2D versions. The initial density profile now becomes:

\begin{equation*}
    \rho = 1 + 3e^{-80(x^2+y^2+z^2)}.
\end{equation*}

Thus we have a spherical pulse expanding outwards. This is run using both a static and a moving mesh. The domain is defined for $0<x,y,z<0.5$ for the static mesh. For the expanding mesh, we start from an initial domain $0<x,y,z<0.3$, similar to the 1D case. The boundary conditions once again remain similar to the 1D and 2D counterparts. The L1 errors for the static mesh and the moving mesh are seen to have second-order convergence, as seen in Figure \ref{fig:3Dpulse}.

\subsection{2D homologous outflow}

For this test, we consider a fluid that has uniform density and pressure throughout. It expands with a homologous velocity, that is:

\begin{equation}
\label{eq:homologous_exp}
    \mathbf{v} = \mathbf{r}/t
\end{equation}

Due to pressure equilibrium, the velocity will remain constant in time for a fluid element.  Thus, $\mathbf{v} = \mathbf{r}/t$ at all times. Assuming this, one can use the continuity equation in the uniform-density regions:

\begin{equation}
    \partial_t \rho + \nabla \cdot ( \rho \mathbf{v} ) = 0.
\end{equation}

Since $\rho$ is uniform in space, we can take it out of the divergence and plug in $\mathbf{v} = \mathbf{r}/t$. Considering this problem in 2 dimensions gives us the first-order (in time) ODE: 

\begin{equation}
    \partial_t \rho = - \rho / t \nabla \cdot ( \mathbf{r} ) = - 2 \rho / t,
\end{equation}

the solution to which is:

\begin{equation}
    \rho(t) = \rho_0 (t/t_0)^{-2},
\end{equation}

where $\rho_0$ is the value of $\rho$ at the initial time $t_0$. Given the analytic value of density, we can formulate an error measure to check for convergence as follows:

\begin{equation}
    \bar{L}_1 = \frac{\langle\rho\rangle}{\rho_0(t/t_0)^{-2}} - 1,
\end{equation}

where $\langle\rho\rangle$ is the volume-averaged density. The calculations are run between $t=1$ and $t=1000$.

This test is performed with both the static mesh and moving mesh. In both cases, the resolution is $(64)^2$. For the former case, we get $L_1=1.0\times10^{-8}$. In contrast, the $L_1$ error for the moving mesh is below machine precision for double type variables in C ($<10^{-16}$). Thus \sprout\, is capable of maintaining homologous outflows to a very high degree of precision for many orders of magnitude in time. This is further demonstrated by setting up a passive scalar map as shown in Figure \ref{fig:homologous}.

\subsection{3D Sedov-Taylor}

\begin{figure}
\centering
\fig{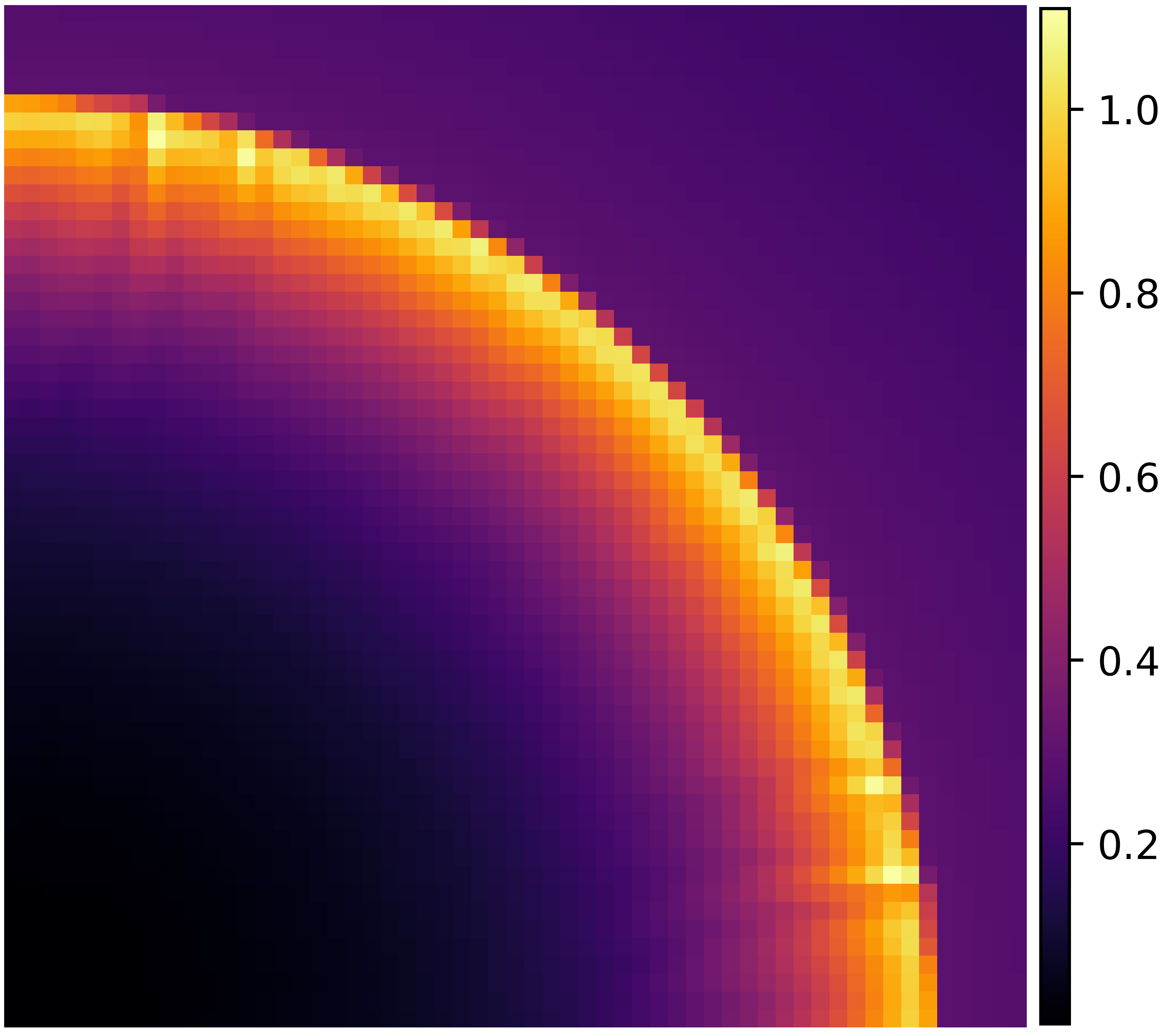}{0.48\textwidth}{}
\caption{Slice in the x-z plane of the density map for the Sedov-Taylor blast wave at $t=10$.  Minor numerical artifacts seen are primarily due to the fact that this test is performed at extremely low resolution $(64 \times 64 \times 64)$.  The hydrodynamic evolution of bulk quantities is captured accurately even at this low resolution, as shown in Figure \ref{fig:Sedov}.
\label{fig:Sedov_slice}}
\end{figure}

\begin{figure}
\centering
\fig{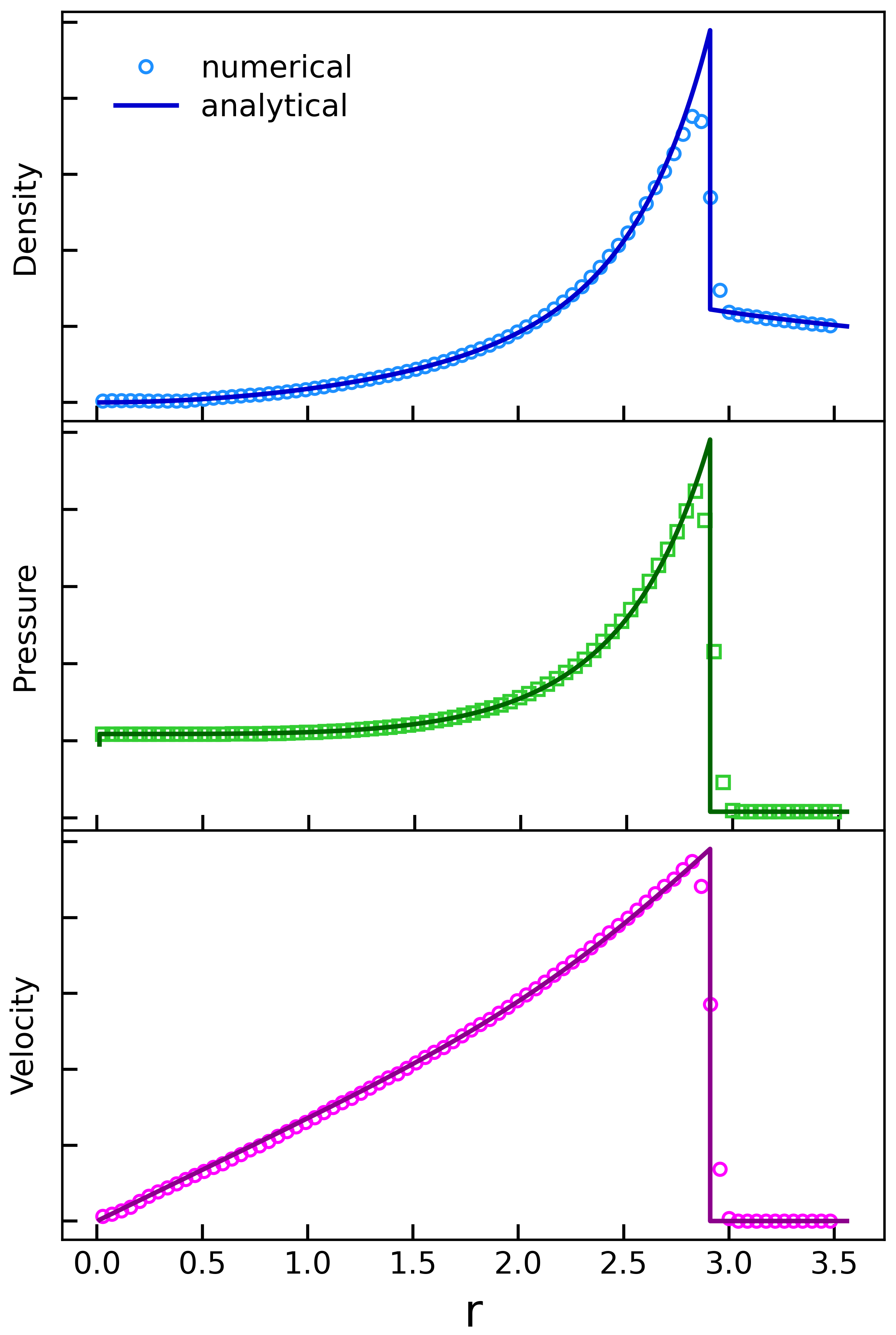}{0.45\textwidth}{}
\caption{Angle-averaged density, pressure, and velocity vs radial distance for the Sedov-Taylor blast wave at $t=10$. \sprout\, data (markers) is compared to the analytic (solid line) curve. The shock propagates against a $\rho \propto r^{-1}$ density profile.  This test employed an extremely low resolution of ($64 \times 64 \times 64$).
\label{fig:Sedov}}
\end{figure}

The Sedov-Taylor explosion problem is used as another test for capturing shocks with non-Cartesian geometry accurately. For this test a large amount of thermal energy is injected within a small region into some ambient medium. A strong shock is formed which advances self-similarly through the medium according to the Sedov-Taylor solution. The shock is stable to fluid instabilities. Thus we expect the self-similar solution to hold throughout several orders of magnitude in time as the fluid expands. The initial conditions for this test are as follows:

\begin{equation}
    P=
    \begin{cases}
    E(\gamma-1)/(4\pi r_0^3/3), & r<r_0 \\
    10^{-4}, & r>r_0
    \end{cases}
\end{equation}

\begin{equation}
    \mathbf{v} = 0,
\end{equation}

\begin{equation}
    \rho = r^{-1},
\end{equation}

where $r^2=x^2+y^2+z^2$ and E is the total energy injected. We set $E=1$ and $r_0=0.01$. This test is performed with a moving mesh to resolve the explosion well at all times. The domain is initially defined for $0<x,y,z<2.5\times10^{-2}$ and has a very low resolution of $(64 \times 64 \times 64)$. The mesh expansion is centered on the origin ($x_0=y_0=z_0=0$). The calculations are performed till $t=10$. An azimuthal slice of the density map (in the $x-z$ plane) of the solution is shown in Figure \ref{fig:Sedov_slice}. Figure \ref{fig:Sedov} shows the spherically averaged density, radial velocity and pressure profiles of our solution, compared against the analytical solution.  The analytic solution was generated using the sedov.tbz\footnote[1]{available at \url{https://cococubed.com/research_pages/sedov.shtml}} code \citep{Kamm+2007}.  \sprout\, is able to capture the shock structure, even at this very low resolution.

\begin{figure*}
\centering
\fig{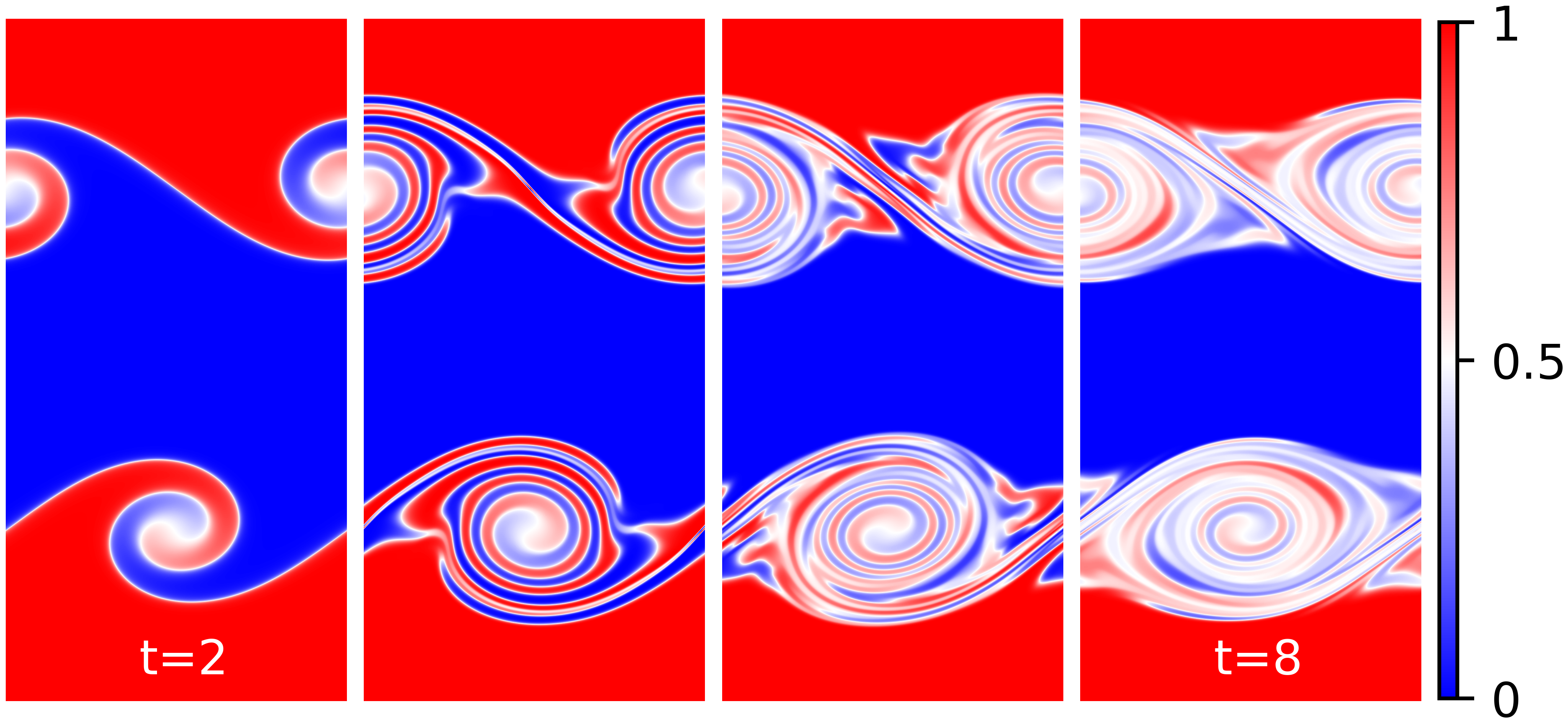}{0.8\textwidth}{}
\caption{Passive scalar map for Kelvin-Helmholtz Instability on a fixed $2048\times2048$ mesh, plotted at regular intervals between $t=2-8$.
\label{fig:KHI}}
\end{figure*}

\begin{figure}
\hspace*{-1cm}
\fig{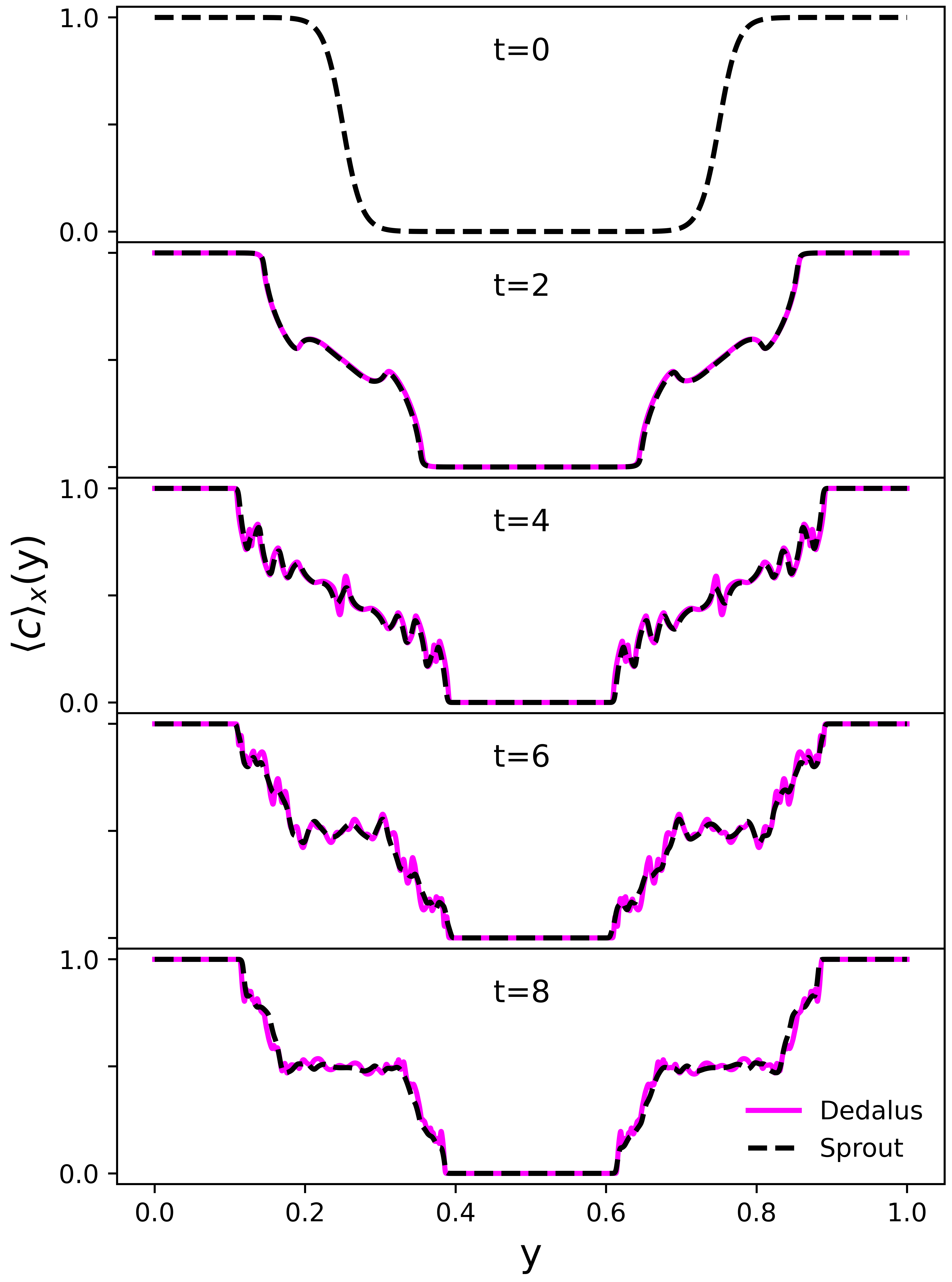}{0.45\textwidth}{}
\caption{Comparison of the quantity $\langle C \rangle _x (y)$ (as defined in (\ref{eq:mean_c1d})) for Kelvin-Helmholtz Instability solutions obtained by \sprout\, to the converged solutions obtained by Dedalus, a pseudo-spectral code \citep{Burns+2020PhRvR}. $\langle C \rangle _x (y)$, or the passive scalar averaged along the flow direction, represents a measure of the amount of mixing for a given solution. The curves from \sprout\, are seen to follow that due to Dedalus to a high degree of accuracy. 
\label{fig:KHI_c1d}}
\end{figure}

\subsection{Kelvin-Helmholtz Instability}

Fluid instabilities provide an effective way to test the artificial diffusion arising in numerical schemes. The first one we study is the Kelvin-Helmholtz instability, which occurs when distinct fluid layers exert a shear force on each other. The initial conditions used here are identical to those for one of the tests proposed by \cite{Lecoanet+2016MNRAS}, to enable a direct comparison between the two solutions. They are as follows:

\begin{equation}
\begin{aligned}
        \rho &= 1 + \frac{\Delta\rho}{\rho_0}\,\frac{1}{2}\left[ \mathrm{tanh}\left(\frac{y-y_1}{a}\right) - \mathrm{tanh}\left(\frac{y-y_2}{a}\right)\right], \\
        P &= 10, \\
        u_x &= \mathrm{tanh}\left(\frac{y-y_1}{a}\right) - \mathrm{tanh}\left(\frac{y-y_2}{a}\right) - 1, \\
        u_y &= A\,\mathrm{sin}(2\pi x)\left[ \mathrm{exp}\left(-\frac{(y-y_1)^2}{\sigma^2}\right) - \mathrm{exp}\left(-\frac{(y-y_2)^2}{\sigma^2}\right) \right], \\
        C &= \frac{1}{2}\left[ \mathrm{tanh}\left(\frac{y-y_1}{a}\right) - \mathrm{tanh}\left(\frac{y-y_2}{a}\right) + 2 \right],
\end{aligned}
\end{equation}

where $A=0.01$, $\sigma=0.2$, $a=0.2$, $y_1 = 0.5$ and $y_2 = 1.5$. We choose $\Delta\rho/\rho_0=0$. The domain is static and defined on $x\in[0,1]$, $y\in[0,2]$, with periodic boundaries. Thus, the middle half of the domain ($y_1<y<y_2$) and the rest of the domain initially contain two distinct fluid layers, with the passive scalar $C=0$ in the former and $C=1$ in the latter. Shear flow is ensured by opposing directions of the x-component of the velocity in the two layers. Fluid density is same for both layers. Snapshots of the passive scalar $C$ is shown in Figure \ref{fig:KHI}, which demonstrates the mixing due to Kelvin-Helmholtz Instability. In addition, we compute the mean of the passive scalar along the direction of flow:

\begin{equation}
\label{eq:mean_c1d}
    \langle C \rangle _x (y) = \frac{1}{L_x}\int C dx,
\end{equation}

where $L_y$ is the length of the domain in the y-direction. \cite{Lecoanet+2016MNRAS} provide resolved, converged numerical solutions to the above initial conditions using the pseudo-spectral code DEDALUS \citep{Burns+2020PhRvR}. We also compute $\langle C \rangle _x (y)$ for the DEDALUS solutions, allowing a more direct comparison of the mixing in \sprout\, to that in the converged solutions. The comparison is shown in Figure \ref{fig:KHI_c1d}, showing that \sprout\ and DEDALUS find agreement on this test.

\begin{figure*}
\centering
\fig{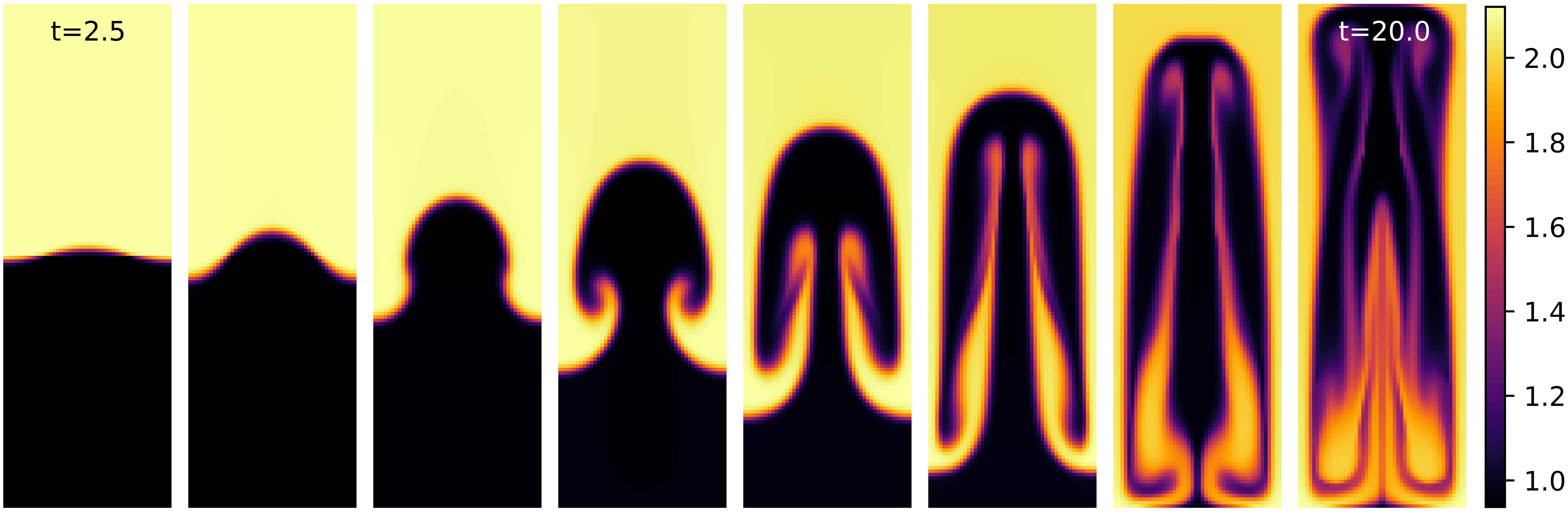}{0.95\textwidth}{}
\caption{Density maps for Rayleigh–Taylor instability on a fixed $48\times144$ mesh, plotted at regular intervals between $t = 2.5-20$. Qualitative agreement is seen between our results and the corresponding snapshots for a similar test done using the fixed mesh configuration of the AREPO code \citep{Springel2010MNRAS}.
\label{fig:RTI}}
\end{figure*}

\begin{figure*}
\centering
\fig{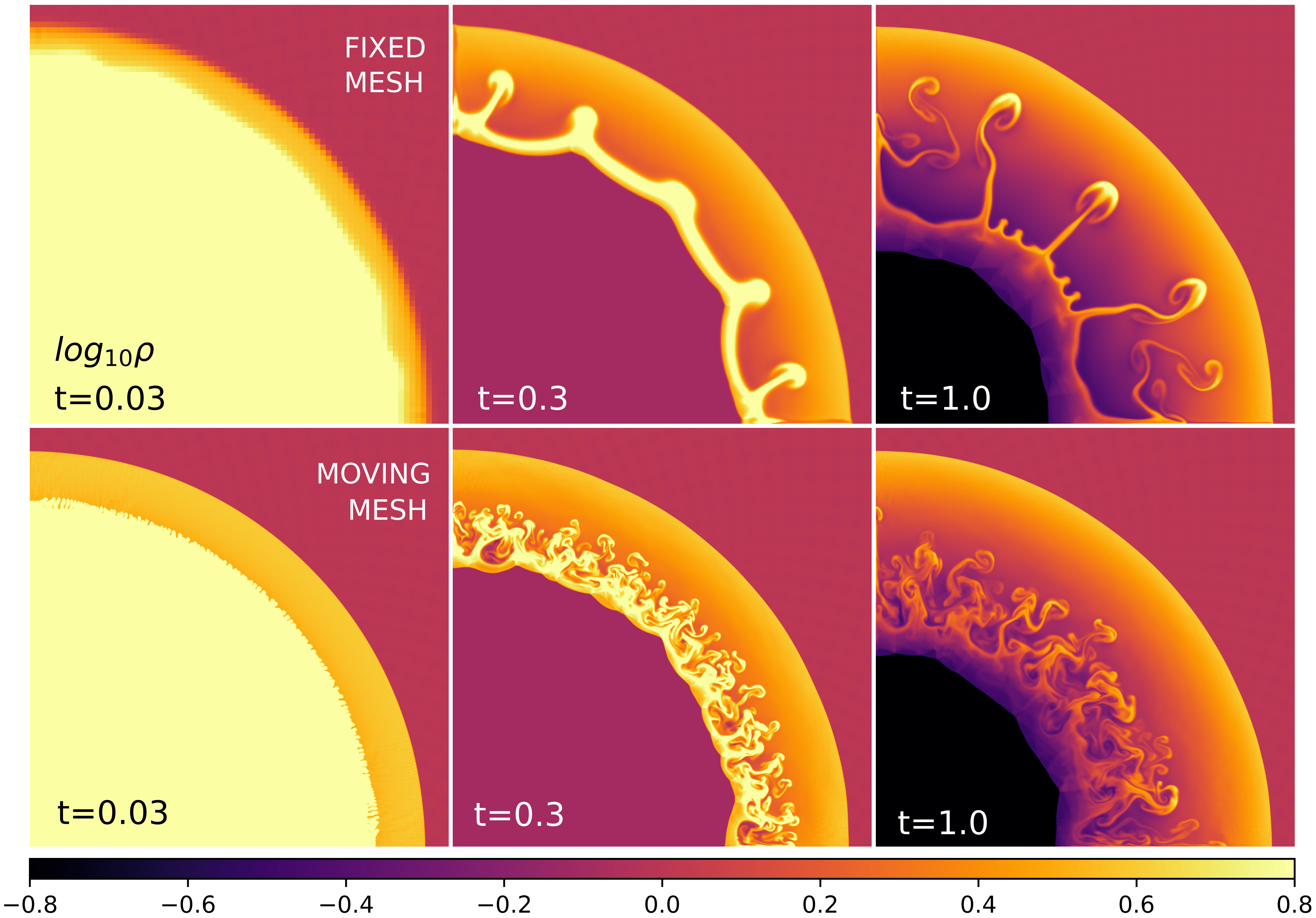}{0.96\textwidth}{}
\caption{Rayleigh-Taylor Instability generated by an explosion with cylindrical symmetry. The upper and the lower panels show logarithm of density from solutions obtained using a fixed and an expanding mesh (both with a resolution of $1000\times1000$), respectively. Each row shows solutions at three  instants of time ($t=0.03,0.3,1.0$) to showcase the evolution of mixing between two fluid layers. All density maps were plotted by zooming onto the expanding blastwave. The effect of this is displayed most drastically by the leftmost image in the top panel, where the blastwave is still relatively small compared to the grid and is resolved poorly. The poor resolution at early times for the fixed mesh discourages mixing at those times, which may be taken to be a numerical artifact. This can be seen by comparing early instants ($t=0.03,0.3$) for the fixed and the expanding mesh configurations.
\label{fig:ERTI}}
\end{figure*}

\subsection{Rayleigh-Taylor Instability in a box}

Another important instance of fluid instability is the Rayleigh-Taylor instability. This involves an unstable equilibrium maintained, for example, by a heavier fluid resting on top of a lighter fluid in a gravitational field. A perturbation of the interface between these two causes growth of Rayleigh-Taylor unstable structures. We carry out this test in the 2D computational domain $(x,y) \in [0,0.5]\times[0,1.5]$ with periodic boundaries, and the following initial conditions:

\begin{equation}
    \rho=
    \begin{cases}
    1.0, & y\leq0.75 \\
    2.0, & y>0.75
    \end{cases}
\end{equation}

\begin{equation}
    P = P_0 + \rho g(y-0.75)
\end{equation}

A perturbation corresponding to a single mode of the Rayleigh-Taylor Instability is added to the y-velocity as follows:

\begin{equation}
    v_y = w_0\left[1-\rm{cos}(4\pi x)\right]\left[1-\rm{cos}(4\pi y/3)\right]
\end{equation}

A gravitational field is added, with the acceleration $\mathbf{g} = g\mathbf{\Hat{y}}$. We choose $P_0 = 5.0$, $w_0=0.0025$, $g=-0.1$, along with $\gamma=1.4$ for this test. The results are shown in Figure \ref{fig:RTI}. Our initial conditions were chosen to be very similar to those for the AREPO code tests \citep{Springel2010MNRAS} to enable a visual comparison between the images.  \sprout\, is able to evolve these initial conditions at an accuracy comparable to a standard fixed-mesh finite volume code; the moving mesh capability was not employed in this test.

\subsection{Cylindrical Rayleigh-Taylor Instability} \label{sec:erti}

Rayleigh-Taylor instabilities also occur in fluid outflows whenever density gradients and pressure gradients with opposite signs develop. This is seen, for example, in young supernova remnants, where expanding stellar ejecta decelerates against the much lighter ambient medium \citep{Chevalier+1992ApJ}. We perform a simplified version of this in 2D, where a cylindrical region of uniform density gas expands homologously against an ambient medium, also of uniform density but lighter:

\begin{equation}
    \rho=
    \begin{cases}
    1/(\pi R^2), & r<R \\
    1, & r>R
    \end{cases}
\end{equation}

\begin{equation}
    \mathbf{v}=
    \begin{cases}
    v_0(\mathbf{r}/R), & r<R \\
    0, & r>R
    \end{cases}
\end{equation}

\begin{equation}
    P = 10^{-5}\rho v_0^2,
\end{equation}

defining $r^2=x^2+y^2$ here, along with setting $R=0.02$ and $v_0=2$. Rayleigh-Taylor instability is seeded by the grid-scale perturbations provided by the Cartesian mesh. The test is performed using both the static mesh and the expanding mesh configurations. \add{For the latter, we do not use an analytic function for the Hubble constant-like quantity H(t) (eq. \ref{eq:mesh_motion}) to set the mesh motion. Instead, we use an algorithm at every timestep to identify the zones housing the shock and select the zone whose fluid element would reach a domain boundary earliest, assuming all elements retain their present velocity. If this zone is too close to the boundary, the value of H at that timestep is chosen depending on the position and fluid speed in this zone so as to keep the shock as stationary as possible with respect to the numerical grid. Thus H(t) isn't given an analytic form before runtime in this case but is calculated in real time.} 

Figure \ref{fig:ERTI} shows the density map at three instants of time ($t=0.03, 0.3, \mathrm{and\,} 1$) for both runs. It is seen that the fixed mesh solutions exhibit little mixing between the expanding fluid and the ambient medium at early times. In fact, the $t=0.3$ snapshot for the fixed mesh solution shows mixing and the onset of turbulent structure formation only at one mode. This is absent for the moving mesh counterpart, where more mixing is seen even at early times, and at many modes. Thus we find the moving mesh to generate the more accurate solution, since in nature, turbulent mixing isn't expected to favor late times or some particular mode.

\section{Summary} \label{sec:summary}

In this work, we present \sprout, a new Cartesian-grid based moving-mesh code. \sprout\, can be found at \url{https://github.com/smandal97/Sprout} under the MIT license \add{or at Zenodo \citep{sprout_zenodo}.}

Numerous code tests have been performed to ensure accuracy and convergence for the novel expanding mesh integrating scheme. \sprout\, captures shocks well, both aligned and misaligned with grid, and ensures second-order convergence for smooth flows. In particular, \sprout\, can capture expanding shocks very well even with a modest resolution when mesh motion is utilized, as demonstrated with the Sedov problem. The expanding mesh capability of \sprout\, is seen to reduce numerical diffusion drastically for outflows, especially when the analytic nature of the bulk flow is known beforehand. The ability to resolve fine fluid structures at small length scales and expand the mesh gradually as the structures grow makes \sprout\, an ideal code to study fluid instabilities in expanding flows like SN explosions and jets. Presently \sprout\, only has the capability to integrate Euler's equations for ideal fluids, but the simple Cartesian mesh provides a very easy way to increase accuracy using higher-order methods, as well as incorporate additional physics like viscous hydrodynamics, self-gravity, magnetohydrodynamics (MHD), and relativistic hydrodynamics. The last two upgrades are being implemented and will be presented in a future work.

\acknowledgments

We thank James Stone and the anonymous referee for helpful comments. We are grateful to Jared E. Bland for suggesting the name Sprout. Numerical calculations were performed on the Petunia computing cluster hosted by the Department of Physics and Astronomy at Purdue University. We thank Chris J. Orr for his extensive help with setting up and debugging Petunia. 

\software{Matplotlib \citep{matplotlib}}

\vspace{15mm}

\bibliographystyle{apj} 
\typeout{}
\bibliography{smbib}

\end{document}